# Link Adaptation Algorithms for Dual Polarization Mobile Satellite Systems

Anxo Tato[1], Pol Henarejos[2], Carlos Mosquera[1], and Ana Pérez-Neira[2,3]

[1] Universidade de Vigo
anxotato@gts.uvigo.es, mosquera@gts.uvigo.es
www.gpsc.uvigo.es
[2] Centre Tecnològic de Telecomunicacions de Catalunya
Castelldefels, Barcelona
pol.henarejos@cttc.es, ana.perez@cttc.cat
www.cttc.cat
[3] Universitat Politècnica de Catalunya, Barcelona

**Abstract.** The use of dual polarization in mobile satellite systems is very promising as a means for increasing the transmission capacity. In this paper we study a system which uses simultaneously two orthogonal polarizations in order to communicate with the users. The application of MIMO signal processing techniques along with Adaptive Coding and Modulation in the forward link can provide remarkable throughput gains up to 100 % when compared with the single polarization system. The gateway is allowed to vary the MIMO and Modulation and Coding Schemes for each frame. The selection is done by means of a link adaptation algorithm which uses a tunable margin to achieve a predefined target Frame Error Rate.

## 1 Introduction

In recent years, the spectrum saturation and the increasing demand for higher data rates in a ubiquitous way encourages the engineers to design new techniques in order to increase the capacity of communication systems without resorting to expand the occupied bandwidth. Two of these techniques are the leverage of multiple antennas at both transmitter and receiver by means of MIMO (Multiple Input Multiple Output) signal processing techniques and also the so-called Adaptive Coding and Modulation (ACM) or link adaptation. Both are part of many current terrestrial wireless communication standards such as LTE [1] and IEEE 802.16 [2] for cellular technologies and IEEE 802.11 [3] for wireless local area networks.

In this paper we propose to apply MIMO techniques to mobile satellite communication systems by exploiting the polarization domain. Satellite links operating at low frequency bands such as L- and S-bands, usually rely on a single circular polarization, either RHCP (Right Hand Circular Polarization) or LHCP (Left Hand Circular Polarization), to avoid the effects of the Faraday rotation [4]. Here we propose the simultaneous use of both orthogonal polarizations within the same beam to communicate with the users following up some preliminary studies presented in [5] within the framework of SatNEx-IV, the European Satellite Network of Experts. An algorithm to switch among different MIMO schemes will be presented. In particular, SISO, Orthogonal Polarization-Time Block Codes (OPTBC), Polarization Modulation (PMod) or V-BLAST (Vertical Bell Laboratories Layer Space-Time) will be chosen as a function of a set of effective Signal to Noise Ratio (SNR) metrics.

On the other hand, the use of ACM techniques allows the system to adapt its instantaneous rate to the current channel capacity without designing the system for the channel worst case. In this paper, following our previous works [6], [7] and [8], we propose the use of a link adaptation algorithm with an adaptive margin for selecting the Modulation and Coding Scheme of each frame (MODCOD or MCS). This algorithm operates together with the MIMO mode selection scheme to increase the spectral efficiency whilst trying to maintain the Frame Error Rate (FER) at a level specified by the Quality of Service (QoS) parameter of the connection.

The remainder of the article is structured as follows. Section 2 provides a description of the satellite communication system, including the link adaptation algorithms, an overview of the channel model and how the channel series are generated. Section 3 describes the link adaptation algorithms for selecting both MIMO mode and MCS. Section 4 provides the simulation results of the algorithms in a maritime mobile scenario. Lastly, in Section 5 we collect the main conclusions and note the future work on this topic.

## 2 System model

We consider a satellite communications system which serves mobile users operating in L- or S-band. A link adaptation algorithm is proposed for the forward link from the gateway towards the mobile terminal (MT). Typically, at these bands only one polarization is used to protect from high cross-polarization leakage. For example, the service BGAN (Broadband Global Area Network), standardized by the ETSI as TS 102 744, only employs RHCP [9]. In higher frequency bands, like Ku and Ka, the high XPD (cross polar discrimination) would make it possible to use two independent streams in both polarizations. The system we propose here, and for which we develop the adaptive algorithms, uses simultaneously both polarizations, yielding a 2x2 MIMO system.

The system that we analyse is narrow-band and single-carrier, and to some extent it is inspired by BGAN. The baseline Single Input Single Output (SISO) system is complemented by additional MIMO modes, namely OPTBC, Polarization Modulation (PMod) and V-BLAST:

– OPTBC (Orthogonal Polarization-Time Block Code) is based on Alamouti space-time coding, used for achieving transmit diversity and introduced in [10]. The spatial components in Alamouti are replaced by the two available polarizations in OPTBC.
- PMod is analogous to Spatial Modulation (SM) techniques [11], with only one polarization per channel use transmitting the symbols from a given constellation. One extra bit of information is conveyed indicating which polarization is chosen. For a QPSK constellation this implies a 50 % gain in spectral efficiency.
- In V-BLAST [12] two independent symbols are transmitted per channel use and the receiver is in charge of reducing the cross-stream interference to perform the detection.

All these three MIMO schemes do not require Channel State Information (CSI) at the transmitter and therefore are suitable for the satellite scenario where the CSIT is outdated due to the long round trip time.

In addition to the forward link, where our algorithms operate, we assume that there is a feedback channel in the return link used by the Mobile Terminal (MT) to inform the gateway about the CSI, the result of the frames decoding (in the form of ACK/NACK) and the optimum transmission mode (which is calculated by the receiver). All these information is used by the gateway, our transmitter, in the selection of the preferred MCS by the link adaptation algorithm we propose.

The signal model for a given time instant $n$ is

$$\boldsymbol{y}_n = \sqrt{P}\boldsymbol{H}_n\boldsymbol{x}_n + \boldsymbol{w}_n \qquad (1)$$

where $P$ is the transmitted power, $\boldsymbol{H}_n \in \mathbb{C}^{2x2}$ is the channel matrix, $\boldsymbol{y}_n \in \mathbb{C}^2$ is the vector of the received signal which has a component per each polarization, $\boldsymbol{x}_n \in \mathbb{C}^2$ is the transmitted signal and $\boldsymbol{w}_n \sim \mathcal{CN}(\boldsymbol{0}, \sigma^2 \boldsymbol{I}_2)$ represents the additive complex white Gaussian noise (AWGN). Therefore, the resultant Signal to Noise Ratio is $\text{SNR} = P/\sigma^2$.

The transmitted symbols are grouped into blocks or packets (codewords) which span $N = 2560$ symbols, transmitted with a baud rate of 33600 symbols/s, which gives a frame length of 80 ms. As we use frames of just one block length, hereafter we will employ the term frame for referring to a codeword. In order to speed up the simulations, we do not implement the entire signal processing block chain and thus, we use Physical Layer Abstraction techniques, in particular the Mutual-Information Effective SNR [6]. The comparison of the effective SNR of a received frame with a threshold SNR of the MCS used to transmit that frame allows us to decide if the frame can be decoded (the effective SNR is higher than the corresponding threshold) or not.

The effective SNR of a frame, $\text{SNR}_{eff}$, given the set of the $N$ SNRs of each symbol $\gamma_n$ can be obtained with

$$\text{SNR}_{eff} = \Phi^{-1}\left(\frac{1}{N}\sum_{n=1}^{N}\Phi(\gamma_n)\right), \qquad (2)$$

procedure called SNR compression in [13]. The SNR $\gamma_n$ for each symbol period is calculated for the different MIMO modes as:

– SISO:
$$\gamma_n = \text{SNR}|[\boldsymbol{H}_n]_{11}|^2 \qquad (3)$$

– OPTBC:
$$\gamma_n = \frac{\text{SNR}}{2}\|\boldsymbol{H}_n\|_F^2 \qquad (4)$$

– PMod:
$$\gamma_n = \text{SNR}\,\frac{|[\boldsymbol{H}_n]_{11}|^2 + |[\boldsymbol{H}_n]_{22}|^2}{2} \qquad (5)$$

– V-BLAST (MMSE (Minimum Mean Square Error) receiver):
$$\gamma_{n,k} = \frac{1}{[(\boldsymbol{I}_2 + \frac{\text{SNR}}{2}\boldsymbol{H}_n^H \boldsymbol{H}_n)^{-1}]_{k,k}} - 1 \qquad (6)$$

In the previous equations $[\boldsymbol{H_n}]_{ij}$ denotes the coefficient $(i,j)$ of the channel matrix for time instant $n$, $k$ indicates the number of the stream (1 or 2), $\boldsymbol{I}_2$ is the $2 \times 2$ identity matrix and $\|\boldsymbol{H_n}\|_F$ denotes the Frobenius norm of the matrix. We should say that in the case of V-BLAST the SNR expression depends on the type of receiver. Throughout this work we assume the MMSE receiver is used for V-BLAST reception, given its simplicity and robustness against noise.

Table 1 collects all the available transmission modes and MCS, including the required effective SNR for correct decoding (obtained from the curve of SNR vs Mutual Information) and corresponding spectral efficiency in bps/Hz.

**Table 1.** Coding rate options for the F80T1Q-1B bearer [9], QPSK constellation.

|  | L8 | L7 | L6 | L5 | L4 | L3 | L2 | L1 | R |
|---|---|---|---|---|---|---|---|---|---|
| Coding rate | 0.34 | 0.40 | 0.48 | 0.55 | 0.63 | 0.70 | 0.77 | 0.83 | 0.87 |
| Threshold SNR ($\text{SNR}_{th}$) (dB) | -2.15 | -1.21 | -0.09 | 0.83 | 1.84 | 2.74 | 3.67 | 4.54 | 5.19 |
| Spectral efficiency SISO/OPTBC | 0.68 | 0.80 | 0.96 | 1.10 | 1.26 | 1.40 | 1.54 | 1.66 | 1.74 |
| Spectral efficiency PMod | 1.02 | 1.20 | 1.44 | 1.65 | 1.89 | 2.10 | 2.31 | 2.49 | 2.61 |
| Spectral efficiency V-BLAST | 1.36 | 1.60 | 1.92 | 2.20 | 2.52 | 2.80 | 3.08 | 3.32 | 3.48 |

### 2.1 Channel model

The simulation of the mobile satellite dual polarized channel has been done following the work of [14]. The channel is obtained as the sum of three different components: the Line-of-Sight (LoS), the specular reflected signal and the diffuse components, produced by the scatterers near the MT. LoS and specular components are modelled as Rice random variables whereas the diffuse component, which causes the fast fading, is Rayleigh distributed. These three components are grouped into the following channel matrix:

Link Adaptation Algorithms for Dual Polarization Mobile Satellite Systems    5

$$\boldsymbol{H} = \boldsymbol{\beta} e^{j\phi} \boldsymbol{K}_L + \boldsymbol{\xi} e^{j\phi} \boldsymbol{K}_S + \boldsymbol{D}\boldsymbol{K}_D \qquad (7)$$

$\boldsymbol{K}_L$, $\boldsymbol{K}_S$ and $\boldsymbol{K}_D$ are the $K$-factor matrices which collect the Rice factors of each polarization for the LoS, specular and diffuse components, respectively. The two matrices $\boldsymbol{\beta}$ and $\boldsymbol{\xi}$ are related to how the channel mixes the two polarizations and matrix $\boldsymbol{D}$ entries are complex Gaussian random variables with a given covariance matrix. The parameters for building all these matrices depend on the considered environment and can be found in [14].

### 2.2 Channel generation

Fig. 1 shows the block diagram of the channel time series generator. One important aspect of the generated channel series is the time correlation and the Doppler spread. Assuming the Clarke's model, the coherence time can be approximated by

$$\tau_c = \frac{3\lambda}{4v\sqrt{\pi}}, \qquad (8)$$

being $\lambda$ the wavelength and $v$ the mobile terminal speed. We first generate $Q = \lceil N/(\tau_c f_s) \rceil$ independent realizations of the channel matrices from the uniformly distributed random phases and the complex Gaussian random matrix $\boldsymbol{D}$. Then we make a linear interpolation to obtain the $N$ channel matrices and a low-pass-filter with cut-off frequency equal to the Doppler spread, $D_s = v/\lambda$, is applied. Lastly, the channel matrices are scaled to yield the given average SNR of the simulation.

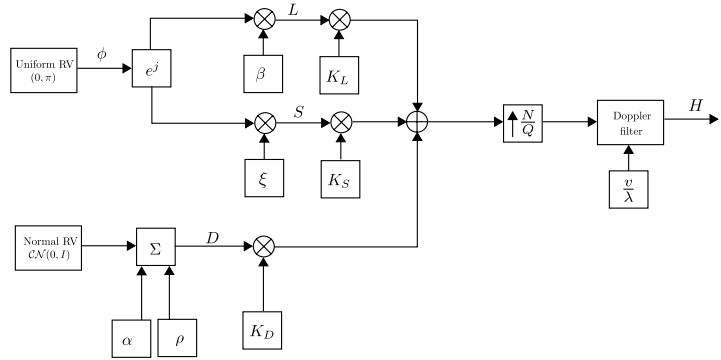

**Fig. 1.** Block diagram of the channel time series generator

## 3 Algorithm for mode and MCS selection

In this paper, the link adaptation procedure has to select the MCS and the MIMO mode (SISO, OPTBC, PMod or V-BLAST). We split these tasks between the gateway and the MT. On the one hand, the MT selects the MIMO mode that achieves the highest throughput given the current SNR. The receiver needs to compute the effective SNR for all modes to choose the most efficient. This task is performed by the receiver to reduce the feedback load. In addition, the transmitter (gateway) is informed by the MT of the ACK/NACK, the preferred MIMO mode and the effective SNR for that mode, and selects consequently the MCS of the following frame.

In order to formalize the procedure let us define the following matrices. $\boldsymbol{S}$ is the matrix with the spectral efficiency of each combination of MIMO mode - MCS. For example, the element $S_{ij}$ is the spectral efficiency of MIMO mode $i$ and MCS number $j$. We also define the vector $\overline{\text{SNR}}_{eff}$, which represents the CSI and whose elements are the effective SNR for each one of the transmission modes. Lastly, we have the vector with the threshold SNR of each MCS $\overline{\text{SNR}}_{th}$; note that this threshold is the same for all MIMO modes.

The optimization problem to choose the optimum mode $\hat{T}$ can be written as

$$\hat{T} = i \quad \text{s. to} \quad R_{ij} = \max_{i,j}\{R_{ij}\}, \tag{9}$$

with the corresponding matrix of rates given by $R_{ij} = S_{ij}[\text{SNR}_{eff,i} \geq \text{SNR}_{th,j}]$. If no MCS verifies the required effective SNR, i.e., all $R_{ij}$ are zero, or in case of a tie, the MIMO mode with the highest effective SNR is chosen and reported back to the gateway.

The choice of the MCS at the gateway follows our previous work in [7], which exploits an adaptive margin. Here, for lack of space, we do not replicate the derivation of the algorithm; those interested readers are referred to [6],[7]. The gateway selects an MCS $m_i$ using a Lookup Table (LUT), represented by means of a function $\Pi(\cdot)$ which maps SNR intervals to MCSs: $m_i = \Pi(\text{SNR}_{i-d} + c_i)$. All this is shown graphically in the diagram of Fig. 2. The value of the SNR reported back by the MT, $\text{SNR}_{i-d}$, plus a margin $c_i$ is introduced in the LUT for selecting the MCS; this margin is updated when new feedback comes in. We assume a round trip time which amounts to the duration of $d$ frames. The recursive equation for updating the value of the margin is

$$c_{i+1} = c_i - \frac{\mu}{\theta^2 + \text{SNR}_{i-2d}^2}(\epsilon_{i-d} - p_0)\theta. \tag{10}$$

This recursion is derived in [6] to solve the following optimization problem

$$\min_c J(c) = \min_c |\mathbb{E}[\epsilon] - p_0|^2. \tag{11}$$

The involved variables are $\text{SNR}_{i-2d}$, the effective SNR used by the LUT to decide the MCS of frame $i-d$, and the acknowledgement $\epsilon_{i-d}$ (1 for ACK, 0 for NACK). $\mu$ and $\theta$ are two constants which take the values 1 and 10, respectively. In this work we suppose the objective FER, $p_0$, has a fixed value, although there are some studies like [15] which show the benefits of having a variable $p_0$ in order to maximize the throughput.

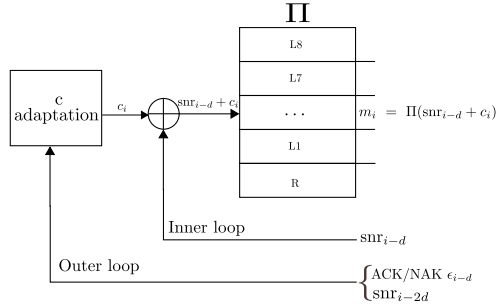

**Fig. 2.** Block diagram of the MCS selection algorithm with the LUT.

## 4 Simulation results

Several simulations are performed to evaluate the spectral efficiency gain from the use of two polarizations simultaneously, and also to understand the potential robustness coming from the use of an adaptive margin in the link adaptation algorithm. All the simulations results presented here are obtained for a maritime scenario with a vessel moving at a constant speed of 50 km/h. The parameters of the channel generator were taken from [14]. The carrier frequency in the simulations is 1.6 GHz, a typical frequency of some Mobile Satellite Systems operating in the L-band. QPSK modulation with frames of 80 ms and 2560 symbols (baud rate of 33600 symbols/s) are used in the physical layer, similarly to the bearer F80T1Q-1B of BGAN [9]. Each simulation comprises the transmission and reception of $N = 60,000$ frames for a specific average SNR, ranging from -5 dB to 25 dB in steps of 2.5 dB. Average spectral efficiency, defined as $(1 - \epsilon_i)r_{m_i}/N$, and cumulative FER during the whole transmission is also computed for each simulation. In the previous expression $r_j$ is the rate of the $j$th MCS, $m_i$ is the MCS selected for frame $i$ and $\epsilon_i$ is the corresponding ACK. Lastly, the RTT is set to 7 frames (560 ms) to approximate the feedback delay for GEO satellites. Contrary to other publications which assume instantaneous feedback, here it has a reasonable value and the proposed scheme is robust even with this delay.

The objective of the first set of simulations is to show the benefits of using two orthogonal polarizations simultaneously to serve a mobile user in low bands of the spectrum. Despite the XPD being lower in these bands than in Ku and Ka bands, it provides a significant throughput gain even with the available power split between the two polarizations. In these simulations the SISO case (transmit only in one polarization) is compared with the use of the two polarizations (MIMO) based on the algorithm for choosing dynamically the MIMO mode of Equation (9). In both cases ACM is employed with a fixed margin $c = -1$ dB in the LUT (see Fig. 2).

The results collected in Table 2 show that use of dual polarization provides a gain higher than 50% for most cases. Interestingly, OPTBC makes it possible to extend the operation range at low SNR below SISO. As the SNR increases the selected mode turns to be PMod instead of OPTBC with gains of up to 55%. At an average SNR of 10 dB both PMod and V-BLAST are employed and for higher SNRs V-BLAST is mainly used. The V-BLAST MIMO mode, which sends two simultaneous streams of symbols, offers up to 100% gain in terms of throughput, i.e., it can double the capacity.

**Table 2.** Results of the comparison of single polarization (SISO) and adaptive dual polarization (MIMO).

| Average SNR | Selected modes | Efficiency (SISO) | Efficiency (MIMO) | Gain |
|---|---|---|---|---|
| -5 dB | OPTBC | 0 bps/Hz | 0.68 bps/Hz | Inf |
| -2.5 to 7.5 dB | PMod | 0.68 - 1.50 bps/Hz | 1.02 - 2.32 bps/Hz | 50 - 55 % |
| 10 dB | PMod (30 %), BLAST (70 %) | 1.69 bps/Hz | 2.47 bps/Hz | 47 % |
| 12.5 to 25 dB | BLAST | 1.73 - 1.74 bps/Hz | 2.77 - 3.48 bps/Hz | 60 - 100 % |

Next we let the margin to adapt in an effort to match a target FER. Figure 3 shows the results obtained in terms of average spectral efficiency and FER. An ACM scheme with a fixed margin of -1 dB is compared with a link adaptation algorithm with adaptive margin for two different values of target FER, 10% and 1%.

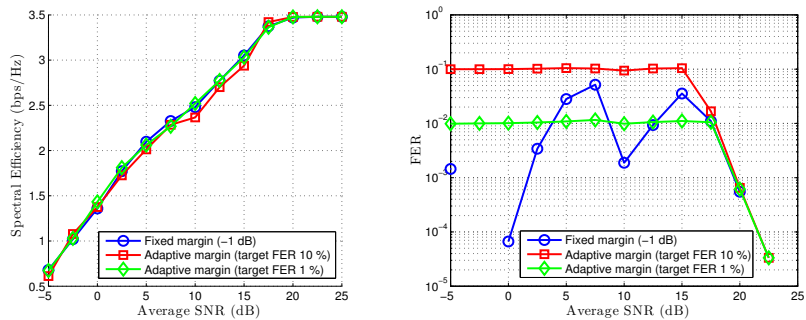

**Fig. 3.** Average spectral efficiency (left) and FER (right)

In Fig. 3 we observe the behaviour of the algorithm in terms of spectral efficiency and FER for a wide range of LOS SNRs. The spectral efficiency grows linearly with the SNR in dB until it reaches its maximum value of 3.48 bps/Hz. When looking only at the throughput there are not significant differences between using a fixed and an adaptive margin. Furthermore, we can observe how the target FER $p_0$ influences the final spectral efficiency. For example, the simulations with the target FER 1 % have a slightly better efficiency than those with target FER 10 %. This dependence of the spectral efficiency with the target error rate was already studied in [15] where the authors show that there is an optimal FER which maximizes the throughput.

In general, the margin $c_i$ converges to a value and oscillates around it, except for the three higher SNRs where it grows indefinitely. We have observed that the value of the final margin depends on the operation point (LOS SNR), the target FER $p_0$, the channel conditions and the mobile speed. This last fact is exemplified in [7] where we show the evolution of the margin in an Intermediate Tree Shadow environment for several MT speeds.

On the other side, the FER behaves very differently when adapting the margin with respect to the fixed case; as shown on the right plot in Figure 3, a fixed margin cannot guarantee a prescribed FER for different SNR values. As opposed to this, the adaptive algorithm exposed earlier matches the target FER, 0.1 and 0.01 in the example, for a wide range of LOS SNRs. Therefore, the combination of the transmission mode selection and the adaptive margin achieves remarkable throughput gains when compared with the SISO case, whilst guaranteeing a physical layer FER suited to the prescribed QoS.

## 5 Conclusions and future work

The simultaneous use of dual orthogonal circular polarizations in L/S-bands for mobile satellite communication systems is proposed in this paper. The application of MIMO techniques to this satellite scenario, by switching among several MIMO modes, generates significant increments in the spectral efficiency that can get close to 100% for high values of SNR, without using extra bandwidth or power. Moreover, we have proposed link adaptation algorithms to select the MIMO mode and the MCS of each frame based on an adaptive margin. This helps the system to adapt itself to the channel conditions and guarantee a prefixed target Frame Error Rate for a wide range of SNRs. Lastly, in the future we plan to improve the algorithm through the online calculation of the optimal FER which maximizes the spectral efficiency.

## 6 Acknowledgements

This work was partially funded by the Agencia Estatal de Investigación (Spain) and the European Regional Development Fund under project MYRADA (TEC2016-75103-C2-2-R). Also funded by the Xunta de Galicia (Secretaría Xeral de Universidades) under a predoctoral scholarship (co-funded by the European Social Fund) and under Agrupación Estratéxica Consolidada de Galicia accreditation 2016-2019 (co-funded by the European Regional Development Fund - ERDF). Part of the research was done during a stay at the Centre Tecnològic de Telecomunicacions de Catalunya (CTTC) supported by the project SatNEx-IV, co-funded by the European Space Agency (ESA). This work has also received funding from the Spanish Ministry of Economy and Competitiveness (Ministerio de Economía y Competitividad) under project TEC2014-59255-C3-1-R and from the Catalan Government (2014SGR1567).